\documentclass[twocolumn, aps, superscriptaddress, prl]{revtex4}

\usepackage{amssymb,amsmath}
\usepackage{graphicx, epsfig}

\def \hf{\tfrac{1}{2}}    

\def \ord{\mathcal{O}}

\def\lba{\left(}    \def\rba{\right)}
\def\lbc{\left[}    \def\rbc{\right]}

\newcommand{\ket}[1]{\left|{#1}\right.\rangle}

\DeclareMathOperator{\tr}{tr}

\begin{document}

\title{Particle partitioning entanglement in itinerant many-particle systems}

\author{O.~S.~Zozulya}
\affiliation{Institute for Theoretical Physics, University of
Amsterdam, Valckenierstraat 65, 1018 XE Amsterdam, the Netherlands}

\author{Masudul Haque}
\affiliation{Max-Planck Institute for the Physics of Complex
Systems, N\"othnitzer Str.~38, 01187 Dresden, Germany}

\author{K.~Schoutens}
\affiliation{Institute for Theoretical Physics, University of
Amsterdam, Valckenierstraat 65, 1018 XE Amsterdam, the Netherlands}

\begin{abstract}

For itinerant fermionic and bosonic systems, we study `particle entanglement',
defined as the entanglement between two subsets of particles making up the
system.
We formulate the general structure of particle entanglement in many-fermion
ground states, analogous to the `area law' for the more usually studied
entanglement between spatial regions.
Basic properties of particle entanglement are uncovered by considering
relatively simple itinerant models.

\end{abstract}


%


\maketitle

\emph{Introduction} ---
In recent years, concepts from quantum information have proved
useful for condensed matter systems.  One prominent example is the
study of the entanglement between a part ($A$) and the rest ($B$)
of the many-particle system, measured by the entanglement entropy
$S_A$.  The entanglement entropy, $S_A = -
\tr\lbc\rho_A\ln\rho_A\rbc$, is defined in terms of the reduced
density matrix $\rho_A = \tr_B\rho$ obtained by tracing out $B$
degrees of freedom.

To define a bipartite entanglement, one has to first specify the partitioning
of the system into $A$ and $B$.  The most commonly used scheme is to partition
space, \emph{e.g.}, partition the lattice sites into $A$ sites and $B$ sites.
However, for itinerant particles, with the wavefunction expressed in
first-quantized form, one can meaningfully partition \emph{particles} rather
than \emph{space}, and calculate entanglements between subsets of particles.
Since each particle has a label in first-quantized wavefunctions,
indistinguishability does not preclude well-defined subsets of particles.
Note that, with such partitioning, $A$ or $B$ do not correspond to connected
regions of space.

The distinction between particle entanglement and spatial entanglement
was made relatively recently \cite{pcle-ent_previous, our-prl-07,
  latorre}.  In work reported since then, particle entanglement has
been shown to be a promising novel measure of correlations
\cite{our-prl-07, latorre, our-prb-07, SantachiaraCabra_JStatMech07,
  KatsuraHatsuda_CalogeroSutherland_JPA07}.  In fractional quantum
Hall states this type of entanglement reveals the \emph{exclusion
  statistics} inherent in excitations over such states
\cite{our-prl-07, our-prb-07}.  Similar insight arises from particle
entanglement calculations in the Calogero-Sutherland model
\cite{KatsuraHatsuda_CalogeroSutherland_JPA07}.  For one-dimensional
anyon states, particle entanglement is found to be sensitive to the
anyon statistics parameter \cite{SantachiaraCabra_JStatMech07,
  SantachiaraCalabrese_anyon}.

Clearly, entanglement between particles in itinerant systems is a promising
new concept, potentially useful for describing subtle correlations and the
interplay between statistics and interaction effects.  A broad study of the
concept and its utility is obviously necessary.
Unfortunately, particle entanglement has till now been studied mostly in
relatively exotic models, so that the literature lacks simple intuition about
these quantities.  This Letter fills this gap.  We provide results for the
simplest nontrivial itinerant fermionic and bosonic models, and present
generic behaviors by generalizing available results.

We first present upper and lower bounds for the entropy of entanglement $S_n$
between a subset of $n$ particles and the remaining $N-n$ particles.  We
formulate a `canonical' asymptotic form for fermionic and some bosonic
systems. We next present results for a two-site Bose-Hubbard model.  Through
this toy model, we identify two general mechanisms of obtaining nonzero
particle entanglement in many-particle models.  One mechanism is simply that
of (anti-)symmetrization of wavefunctions, while the other is due to the
formation of `Schr\"odinger cat'-like states.  The second mechanism is shown
to be fragile, in the same sense that cat states are fragile in macroscopic
settings.
We next switch to true lattice models, focusing on spinless fermions on a
one-dimensional (1D) lattice with nearest-neighbor repulsion, sometimes known
as the $t$-$V$ model.  We find similar mechanisms at work, in a nontrivial
setting.  In addition, our study of the $t$-$V$ model enables us to present
generic intuition about particle entanglement in many-particle systems,
expressed in our canonical asymptotic language.

\emph{Bounds} --- A generic itinerant lattice system has $N$ particles
in $L$ sites; we consider bosons or spinless fermions so that
$N\leq{L}$.
In every case, a natural upper bound for $S_n$ is provided by the (logarithm
of the) size of the reduced density matrix $\rho_A = \rho_n$, \emph{i.e.}, the
dimensions of the reduced Hilbert space of the $A$ partition.  This size is
$\begin{pmatrix}L\\ n\end{pmatrix} = C(L, n)$
for fermions and $C(L-1+n, n)$ for bosons.  The actual rank of $\rho_n$ can be
much smaller due to physical reasons, so that the entanglement entropies
are usually significantly smaller than the upper bounds, as we shall see in
the examples we treat.

In a bosonic system, $S_n$ can vanish, since a Bose condensate wavefunction is
simply a product state of individual boson wavefunctions, each identical.  For
fermions, however, anti-symmetrization requires the superposition of product
states; for free fermions this causes $\rho_n$ to have $C(N,n)$ equal
eigenvalues.  This provides a nonzero lower bound for $S_n$ in a fermionic
system.
\begin{gather}  \label{eq_bounds_b}
{\rm Bosons:} \qquad 0 \leq S_n \leq \ln C(L-1+n, n) \; , \\
\label{eq_bounds_f}
{\rm Fermions:} \qquad \ln C(N,n) \leq S_n \leq \ln C(L, n) \; .
\end{gather}

\emph{Canonical form} ---
For large fermion number, $N{\gg}1$, we propose the following
canonical form for the entanglement of $n{\ll}N$ fermions with the
rest:
\begin{equation} \label{eq_canonical}
\begin{split}
S_n (N) ~&=~  \ln C(N,n) ~+~ \alpha_n ~+~ \ord(1/N) \\ 
&=~ n \ln{N} ~+~ \alpha'_n ~+~  \ord(1/N) \; .
\end{split}
\end{equation}
 This form is suggested by results reported in
Refs.~\cite{our-prl-07,our-prb-07,
KatsuraHatsuda_CalogeroSutherland_JPA07,
SantachiaraCabra_JStatMech07}, and in this Letter.
For example, $\alpha_n = n \ln m$ for the Laughlin state at filling
$\nu=1/m$ \cite{our-prl-07}.
The same canonical behavior seems to hold for bosonic systems which lack
macroscopic condensation into a single mode, \emph{e.g.}, bosonic Laughlin
states \cite{our-prb-07}, or hard-core repulsive bosons in one dimension
\cite{Calabrese_personal}.

Subtle correlation and statistics effects can be contained in the behavior of
the $\ord(1)$ term $\alpha_n$, and sometimes also the $\ord(1/N)$ term.  Our
calculations provide important intuition about how such effects show up in
$\alpha_n$, as we summarize at the end of this Letter.

Note that, for lattice sizes larger than $N$, the generic behavior
\eqref{eq_canonical} indicates that the entanglement entropy does not saturate
the upper bound \eqref{eq_bounds_b} or \eqref{eq_bounds_f} obtained from the
size of the reduced Hilbert space.

\emph{Two-site Bose-Hubbard model} ---
We start with a toy lattice model, with only two sites.  We will consider $N$
bosons on this `lattice', subject to a Bose-Hubbard model Hamiltonian, to
elucidate the basic mechanisms by which an itinerant quantum system can
possess particle entanglement.
The Hamiltonian is
\begin{equation}  \label{eq_BoseHubbard-2site}
\hat{H} ~=~ - \lba \hat{b}_1^{\dagger}\hat{b}_2 +
\hat{b}_2^{\dagger}\hat{b}_1 \rba
~+~ {\hf}U \lba \hat{b}_1^{\dagger} \hat{b}_1^{\dagger} \hat{b}_1 \hat{b}_1
+ \hat{b}_2^{\dagger} \hat{b}_2^{\dagger} \hat{b}_2 \hat{b}_2 \rba   \, .
\end{equation}
For $U=0$, the system is a non-interacting Bose condensate, with each boson
packed into the state $\tfrac{1}{\sqrt{2}}\lba\ket{1}+\ket{2}\rba$.
In the $U\rightarrow +\infty$ case, the system is a Mott insulator, with half
the particles in site 1 and the other half in site 2.  Such a state is simple
in the "site" basis (second-quantized wavefunction), but involves
symmetrization in the "particle" basis (first-quantized wavefunction), leading
to nonzero particle entanglement entropy.
Finally, the $U\rightarrow -\infty$ limit involves all particles in either
site 1 or site 2.  The ground state is a linear combination of these two
possibilities, which for large $N$ is a macroscopic `Schr\"odinger cat' state.
Such a state is somewhat artificial, because an infinitesimal energy imbalance
between the two states will `collapse' this state.
For example, a `symmetry-breaking' term of the form $\epsilon \;
\hat{b}_1^{\dagger} \hat{b}_1$, added to the Hamiltonian
\eqref{eq_BoseHubbard-2site}, would favor site 2 and destroy the cat state.
The resulting state is a product state with zero particle entanglement.

Incidentally, a 2-site model with off-site interaction $V$ (instead of on-site
$U$) has similar physics, with negative (positive) $V$ playing the role of positive
(negative) $U$.

\begin{figure}
\centering
\includegraphics*[width=0.9\columnwidth]{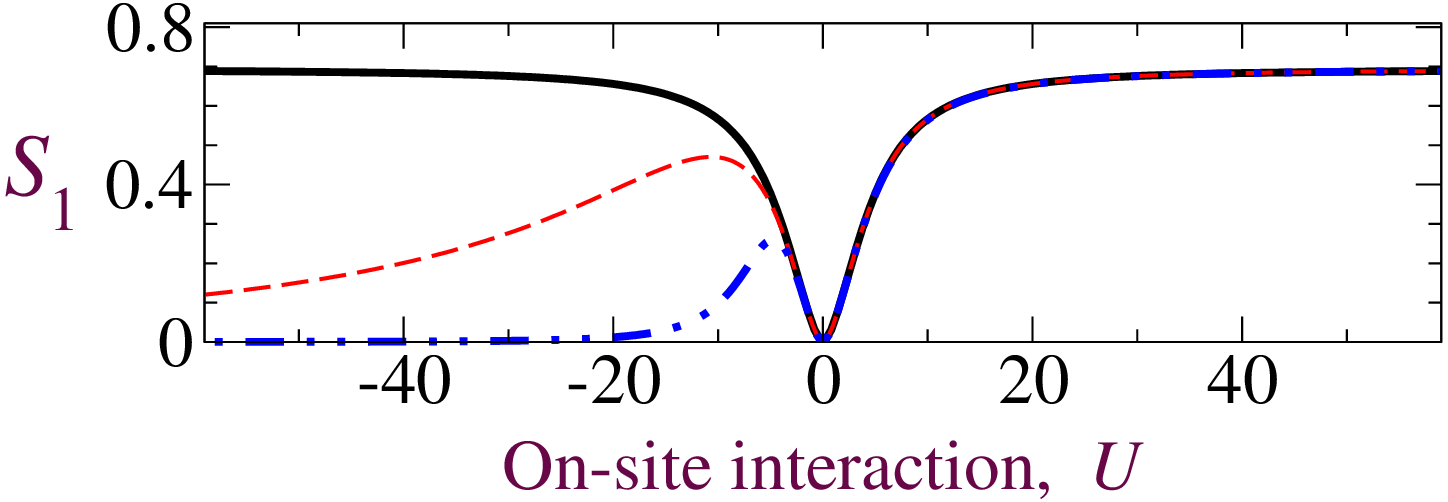}
\caption{ \label{fig_2bosons-2sites}
(Color online.)  Entanglement between two bosons, in the ground state of a
two-site lattice model with on-site repulsion.  Solid curve is for the basic
Bose-Hubbard model.  Dashed ($\epsilon=0.1$) and dash-dotted ($\epsilon=0.1U$)
curves illustrate fragility of `cat' state via a  $\epsilon \hat{b}_1^{\dagger} \hat{b}_1$ term.
}
\end{figure}

\emph{Two bosons in two sites} ---
There is only one way of partitioning two particles ($n=1$), so the only $S_n$
is $S_1$.
We expect $S_1=0$ at $U=0$, and maximal entanglement $S_1=\ln{2}$ for
both `Mott' state at $U={+}\infty$ and the `Schr\"odinger cat' state
at $U={-}\infty$.
The Hilbert space is small; one can diagonalize the problem and
calculate $S_1$ analytically as a function of $U$.  We find $S_1(U) =
S_1(-U)$ interpolating smoothly between zero and $\ln2 \simeq 0.6931$ in
both positive and negative directions (Fig.~\ref{fig_2bosons-2sites}.)

We also demonstrate the fragility of the cat state by showing the effect of an
$\epsilon \hat{b}_1^{\dagger} \hat{b}_1$ term.  There is no appreciable effect
for $U>0$, but for $U<0$ the cat state is destroyed and we get
$S_1\rightarrow\infty$ for $U\rightarrow{-}\infty$.

\begin{figure}
\centering
\includegraphics*[width=0.97\columnwidth]{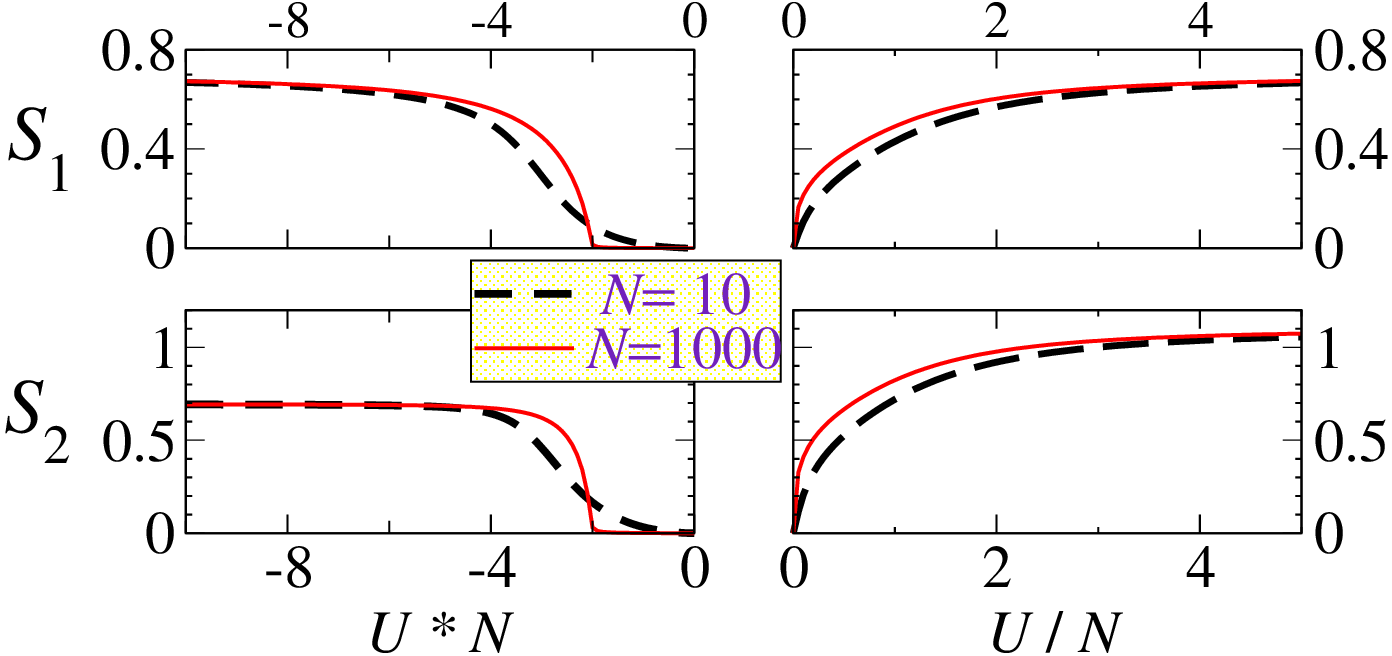}
\caption{ \label{fig_10-1000bosons-2sites}
(Color online.) 1- and 2- particle entanglement entropies for $N$ bosons in two
sites.
}
\end{figure}

\emph{Many bosons in two sites} ---
For $N$ bosons, it is meaningful to study $S_n$ with $n>1$.  Labeling the
basis states by site occupancies, \emph{i.e.}, as $\ket{N_1,N_2}$, the Mott
and cat ground states are respectively $\ket{N/2,N/2}$ and $(\ket{0,N}+
\ket{N,0})/\sqrt{2}$.  The $n$-particle reduced Hilbert space has dimension
$n+1$; the reduced-space basis states can be labeled by the number of $A$
bosons in site 1.  In the Mott state $\ket{N/2,N/2}$, only the diagonal
elements of $\rho_n$ are nonzero and they are all equal; hence
$S_n(U\rightarrow\infty)= \ln(n+1)$.  In the cat state, only two elements are
nonzero, both on the diagonal; hence $S_n(U\rightarrow-\infty)= \ln{2}$,
independent of $n$.
Fig.~\ref{fig_10-1000bosons-2sites} demonstrates, via calculation from
wavefunctions obtained by numerical diagonalization, that $S_n$ increases to
$\ln(n+1)$ and  $\ln{2}$ in the $U\rightarrow\pm\infty$ limits.

Both $\rho_n(U)$ and $S_n(U)$ can be understood in greater detail using
approximations available in the literature
\cite{MuellerHoUedaBaym_fragmentation}.  For $U>0$, the coefficients
$\Psi_{N_1}$ of the ground state $\ket{GS} = \sum_{N_1} \Psi_{N_1}
\ket{N_1,N-N_1}$ can be approximated by a gaussian $\Psi_{N_1} \propto
\exp\lbc(N_1-\hf{N})^2/\sigma^2\rbc$, with $N\sigma^{-2} = (1+U N)^{1/2}$.
The reduced density matrix then has off-diagonal elements of the form
$\exp(-c/\sigma^2)$, which vanish as $U$ increases to the Mott limit.  For
$U<0$ and $|U| N \gtrsim 2$, the function $\Psi_{N_1}$ can be approximated by
two gaussians centered at separate points around $N_1=N/2$.  As the two peaks
sharpen, we converge to the two-eigenvalue case described for $U \rightarrow
-\infty$.  Fig.~\ref{fig_10-1000bosons-2sites} shows that $S_n$ changes rather
sharply around $U\sim{-2/N}$, for large $N$.

To summarize our findings from the bosonic model, we note that the Mott state
for $U>0$ and Schr\"odinger cat state for $U<0$ both possess particle
entanglement.  We have thus identified two generic mechanisms for generating
particle entanglement in itinerant systems.

\emph{Spinless fermions in one dimension} ---
We will now consider the 1D $t$-$V$ model; $N$ spinless fermions
on  $L$ sites with periodic boundary conditions:
$$
H = -t \sum_{<ij>} \lba c^{\dag}_i c_{i+1} +  c^{\dag}_{i+1} c_{i} \rba + V n_i n_{i+1} \; .
$$
We will use $t=1$ units.  For repulsive interactions at half filling
($N=\hf{L}$), this model has a quantum phase transition at $V=2$, from a
Luttinger-liquid phase at small $V$ to a charge density wave (CDW) phase at
large $V$.  We will focus mainly on $V>0$.
This model
is solvable by the Bethe ansatz;  however, calculating particle entanglement
entropies $S_n$ using the Bethe ansatz is a nontrivial problem which we do not
address here.

\emph{Limits} ---
For $V=0$ (free fermions), the ground state is simple in terms of
momentum-space modes: a Slater determinant of the $N$ fermions occupying the
$N$ lowest-energy modes.  The $n$-particle reduced density matrix has
${C(N,n)}$ equal eigenvalues, so that $S_n = \ln\lbc{C(N,n)}\rbc$, independent
of the lattice size $L$.

In the infinite-$V$ limit, the ground state and hence particle entanglement
can be simply understood for the case of half filling, $N={\hf}L$.  The ground
state is an equal superposition of two `crystal' states, and each of them
gives a separate contribution to the reduced density matrix.  The reduced
density matrix has rank $2 C(N,n)$ and equal eigenvalues: $S_n =
\ln\lbc{2C(N,n)}\rbc$.
In the notation of Eq.~\eqref{eq_canonical}, the subleading term $\alpha_n$
intrapolates between $\alpha_n = 0$ at $V = 0$ and $\alpha_n\rightarrow\ln{2}$
for $V\rightarrow\infty$ for half filling. The interpolation details depend on
$n$ and $N$.

\begin{figure}
\centering
 \includegraphics*[width=0.9\columnwidth]{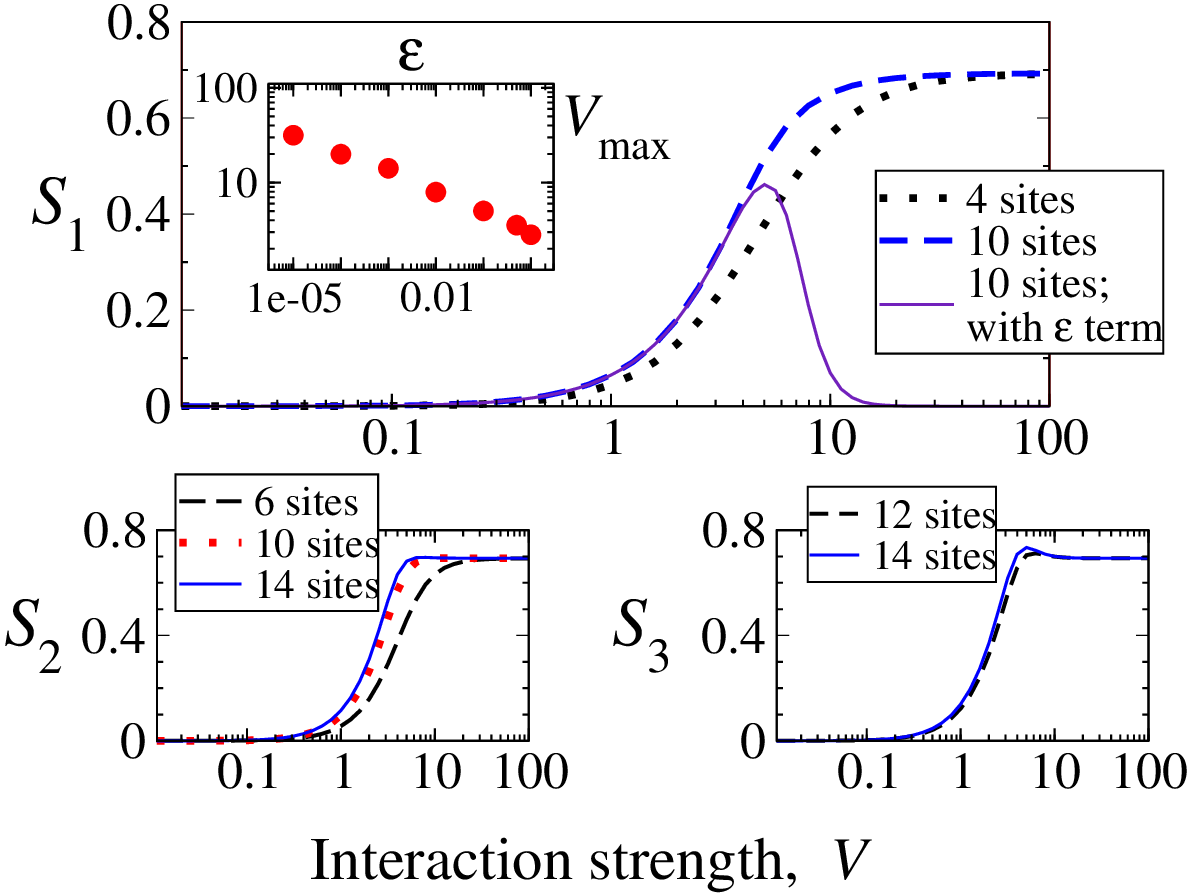}
\caption{ \label{particle-entropies_half-filling}
(Color online.) $n=1$, $n=2$ and $n=3$ entanglement entropy in
  half-filled $t$-$V$ model ($N=L/2$).  The free-fermion contribution
  $\ln\lbc{C(N,n)}\rbc$ has been subtracted off.  The $n=1$ plot also
  displays the effect of a symmetry-breaking
  $\epsilon{c_1^{\dagger}}c_1$ term, with $\epsilon=0.1$.  Inset:
  position of the maximum as function of $\epsilon$.
}
\end{figure}

\emph{Numerical results} ---
For half-filling ($N=\hf{L}$), Fig.~\ref{particle-entropies_half-filling}
presents $S_n(V)$, calculated from wavefunctions obtained by direct numerical
diagonalization.  The $S_n(V)$ function evolves from $S_{\rm
FF}=\ln\lbc{C(N,n)}\rbc$ to $\ln\lbc{2C(N,n)}\rbc \simeq S_{\rm FF}+0.6931$.
For $n>1$, we see non-monotonic behavior in some cases.  At present we have no
detailed understanding of the states or particle entanglements at finite
nonzero $V$.

As in our bosonic model, we see Schr\"odinger cat physics in the $t$-$V$ model
also: the $V=+\infty$ ground state is a superposition of two CDW states of the
form $\ket{101010...10}$ and $\ket{010101...01}$.  The fragility of this cat
state can be seen by adding a single-site potential,
$\epsilon{c}_1^{\dagger}c_1$, or a staggered potential,
$\epsilon'\sum_{i}c_{2i}^{\dagger}c_{2i}$. The ground state then collapses to
a single crystal wavefunction, and $S_n$ drops to $\ln\lbc{C(N,n)}\rbc$
(Fig.~\ref{particle-entropies_half-filling} top panel).

\emph{Phase transition} ---
The particle entanglement entropy shows no strong signature of the
phase transition at $V=2$.  This is also true after extrapolating to
the $N\rightarrow\infty$ limit.  (The extrapolated curves are very
close to the largest-$N$ curves displayed in
Fig.~\ref{particle-entropies_half-filling} and so are not shown.)
The lack of transition signature is not too surprising, because in the
definition of the particle entanglement, the notion of distances (space)
enters rather weakly.  Thus $S_n$ is not sensitive to characteristics of phase
transitions, such as diverging correlation length or large-scale fluctuations.

\begin{figure}
\centering
 \includegraphics*[width=1.0\columnwidth, angle=0]{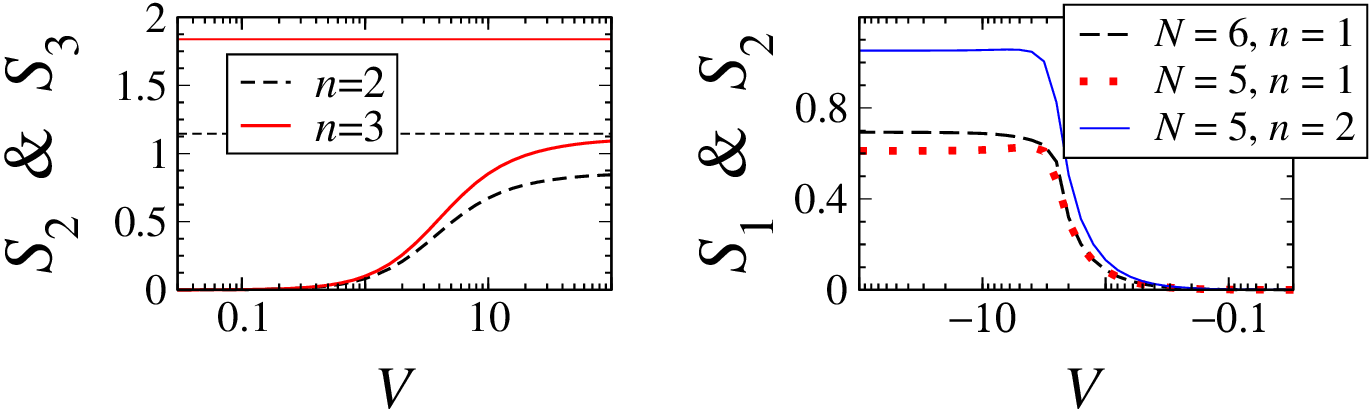}
\caption{ \label{fig_non-hf_negV} (Color online)
(a) $S_n$ for $N=7$, $L=12\neq{2N}$. Horizontal lines are
  corresponding maximal bounds $\ln\lbc{C}(L,n)\rbc$.
(b) negative $V$, half-filling.
Free-fermion contribution $\ln\lbc{C}(N,n)\rbc$ has been subtracted off in
each case.
}
\end{figure}

\emph{Away from half-filling} ---
For $N{\neq}L/2$, the behavior is qualitatively similar to the
half-filled case, $\alpha_n$ increasing from zero to an $\ord(1)$
value as $V$ increases from zero to infinity.
(Fig.~\ref{fig_non-hf_negV}.)  However, there is no simple picture
for the $V\rightarrow\infty$ limit.  Also, $\alpha_n(V)$ appears
to be monotonic, perhaps because $\alpha_n(V\rightarrow\infty)$ is
not constrained as in the half-filled (CDW) case.

Note that, except for $S_{n=1}$ in the half-filled case, the particle
entanglement never saturates the upper bound, $\ln\lbc{C}(L,n)\rbc$,
dictated by Hilbert space size.

\emph{Negative V} ---
An attractive interaction causes the fermions to cluster.  In the
$V\rightarrow -\infty$ limit, the ground state is a superposition (cat
state) of $L$ terms, each a cluster of the $N$ fermions.  The cat
state can be destroyed as in the positive-$V$ case.  For half-filling
with even $N$, the $V\rightarrow -\infty$ wavefunction yields
$S_1=\ln{N}+\ln{2}$.  There are $\ord(N^{-1})$ corrections for odd
$N=L/2$.

\emph{Eigenvalue spectrum (majorization)} ---
The full eigenvalue spectrum of the reduced density matrices
($\rho_n$) of course contain more information than the $S_n$
alone.  For $n=1$ where $S_1(V>0)$ is monotonic, we
numerically observe `majorization' (\emph{e.g.},
Ref.~\cite{Majorization}) of spectra. Obviously, there are many
other aspects of the full spectra that remain unexplored.

\emph{Bosons} ---
An important issue concerns bosonic systems which have partial condensation
into a single mode, so that the leading asymptotic term is not $\ln{N}$.  We
have treated one example, closely related to the fermionic $t$-$V$ model:
hard-core bosons on a 1D lattice (forbidden multiple occupancy, $U=\infty$)
with nearest-neighbor interaction $V$.  The point $V=-2$ has a `simple' ground
state \cite{YangYang66}, which we exploited to find
\[
S_n = \nu n \ln{N} + \ord(N^0)
\]
where $\nu=N/L$ is the filling fraction.  A natural interpretation is that the
pre-factor represents the un-condensed fraction.  Whether this is generic for
bosonic systems with partial condensation remains an intriguing open question.

\emph{Correlations in subleading term} ---
The canonical relation $S_n(N) = \ln\lbc{C(N,n)}\rbc + \alpha_n$ allows us to
formulate correlation effects in terms of the function $\alpha_n$.  For free
fermions, for CDW states of the $t-V$ model, and for Laughlin states
\cite{our-prl-07, our-prb-07}, we have
\[
 \alpha_n({\rm FF}) = 0 \, , \; \alpha_n({\rm CDW}) = \ln{2} \, , \;
 \alpha_n({\rm FQH}) = n\ln{m}  \; .
\]
We note that states which are intuitively `more nontrivially
correlated' have stronger $n$-dependence in $\alpha_n$.  This
strongly suggests that the $\alpha_n$ function is a measure of
correlations in itinerant fermionic states.  It is natural to
conjecture that the linear behavior of $\alpha_n$ is symptomatic
of intricately correlated states like quantum Hall states, and
that in generic itinerant states $\alpha_n$ will have sublinear
dependencies on $n$.

\emph{Equal partitions} ---
In addition to the $n\ll{N}$ behavior we have focused on here, another
promising quantity is $S_{n=N/2}$.  In Ref.~\cite{our-prb-07} we presented
close bounds for this quantity, showing that for fractional quantum Hall
states of given filling $S_{n=N/2}$ tends to be higher for more correlated
states.  For example $S_{n=N/2}$ for a Moore-Read state is higher than that
for a Laughlin state.

\emph{Conclusions} ---
Particle entanglement is  an emerging important measure of
correlations in itinerant many-particle quantum systems.  In this
work, we have set the framework for future studies of the
asymptotic behavior of particle entanglement.  We have also
explored these quantities in relatively simple itinerant models.
We have pointed out several different mechanisms for particle entanglement in
itinerant quantum states, such as localization, Schr\"odinger cat states, and
of course anti-symmetrization of fermionic systems.
Since particle entanglement is a relatively new quantity on which little
intuition is available, these results will form a much-needed basis for future
studies.

Our work opens up a number of questions.  Our considerations have led to an
intriguing speculation for bosonic systems, relating the leading term in the
asymptotic ($N\rightarrow\infty$) expression for $S_n(N)$ to the extent of
Bose condensation.  A thorough study, addressing several bosonic systems, is
clearly necessary.  In the same asymptotic form, one would also like to have a
detailed characterization of how the subleading term $\alpha_n$ describes
correlations.  More concretely, one could ask ``how correlated'' a state needs
to be, in order to have a linear $\alpha_n$ function.


We thank P.~Calabrese and J.~S.~Caux for discussions. M.H.\ thanks
the European Science Foundation (INSTANS programme) for a travel
grant.  O.S.Z.\ and K.S.\ are supported by the Stichting voor
Fundamenteel Onderzoek der Materie (FOM) of the Netherlands.


\begin{thebibliography}{99}

\bibitem{pcle-ent_previous} H.~M.~Wiseman and J.~A.~Vaccaro,
 Phys.\ Rev.\ Lett.\ {\bf 91}, 097902 (2003).
%
M.~R.~Dowling, A.~C.~Doherty, and H.~M.~Wiseman,
  Phys. Rev. A {\bf 73}, 052323 (2006).
%
Y.~Shi, J.~Phys.~A {\bf 37}, 6807 (2004).




\bibitem{our-prl-07}
M.~Haque, O.~Zozulya, and K.~Schoutens, Phys. Rev. Lett.~{\bf 98}, 060401 (2007).

\bibitem{our-prb-07}
O.S.~Zozulya, M.~Haque, K.~Schoutens, and E.H.~Rezayi,
Phys.~Rev.~B~{76}, 125310 (2007).


\bibitem{KatsuraHatsuda_CalogeroSutherland_JPA07}   H.~Katsura and Y.~Hatsuda,
  J.~Phys.\ A: Math.\ Theor.\ {\bf 40}, 13931 (2007).

\bibitem{SantachiaraCabra_JStatMech07}  R.~Santachiara, F.~Stauffer, D.~Cabra,
  J.~Stat.\ Mech.\ L05003 (2007).

\bibitem{latorre}
S.~Iblisdir, J.I.~Latorre, and R.~Orus, Phys. Rev. Lett.~{\bf 98}, 060402 (2007).

\bibitem{SantachiaraCalabrese_anyon} R.~Santachiara and P.~Calabrese,
  preprint, arxiv:0802.1913v1 [cond-mat.mes-hall].

\bibitem{Calabrese_personal} P.~Calabrese, unpublished.

\bibitem{MuellerHoUedaBaym_fragmentation}
E.~J.~Mueller, T.-L.~Ho, M.~Ueda, G.~Baym, Phys.\ Rev.\ A {\bf 74},
033612 (2006).


\bibitem{Majorization} R.~Or\'{u}s, Phys.\ Rev.\ A {\bf71}, 052327 (2005).


\bibitem{YangYang66} C.~N.~Yang and C.~P.~Yang, Phys.\ Rev.\ {\bf 151} 258 (1966).


\end{thebibliography}
\end{document}